\documentclass[aps,showpacs,superscriptaddress,preprint]{revtex4}%
\usepackage{amsfonts}
\usepackage{amsmath}
\usepackage{amssymb}
\usepackage{graphicx}%
\providecommand{\U}[1]{\protect\rule{.1in}{.1in}}

\begin{document}
\title{Ground state properties and high pressure behavior of plutonium dioxide:
Systematic density functional calculations}
\author{Ping Zhang}
\affiliation{LCP, Institute of Applied Physics and Computational Mathematics, Beijing
100088, People's Republic of China}
\author{Bao-Tian Wang}
\affiliation{Institute of Theoretical Physics and Department of Physics, Shanxi University,
Taiyuan 030006, People's Republic of China}
\author{Xian-Geng Zhao}
\affiliation{LCP, Institute of Applied Physics and Computational Mathematics, Beijing
100088, People's Republic of China}


\begin{abstract}
Plutonium dioxide is of high technological importance in nuclear fuel cycle
and is particularly crucial in long-term storage of Pu-based radioactive
waste. Using first-principles density-functional theory, in this paper we
systematically study the structural, electronic, mechanical, thermodynamic
properties, and pressure induced structural transition of PuO$_{2}$. To
properly describe the strong correlation in Pu $5f$ electrons, the local
density approximation$+U$ and the generalized gradient approximation$+U$
theoretical formalisms have been employed. We optimize $U$ parameter in
calculating the total energy, lattice parameters, and bulk modulus at
nonmagnetic, ferromagnetic, and antiferromagnetic configurations for both
ground state fluorite structure and high pressure cotunnite structure. Best
agreement with experiments is obtained by tuning the effective Hubbard
parameter $U$ at around 4 eV within LDA$+U$ approach. After carefully testing
the validity of the ground-state calculation, we further investigate the
bonding nature, elastic constants, various moduli, Debye temperature,
hardness, ideal tensile strength, and phonon dispersion for fluorite PuO$_{2}%
$. Some thermodynamic properties, e.g., Gibbs free energy, volume thermal
expansion, and specific heat, are also calculated. As for cotunnite phase,
besides elastic constants, various moduli, and Debye temperature at 0 GPa, we
have further presented our calculated electronic, structural, and magnetic
properties for PuO$_{2}$ under pressure up to 280 GPa. A metallic transition
at around 133 GPa and an isostructural transition in pressure range of 75-133
GPa are predicted. Additionally, as an illustration on the valency trend and
subsequent effect on the mechanical properties, the calculated results for
other actinide metal dioxides (ThO$_{2}$, UO$_{2}$, and NpO$_{2}$) are also presented.

\end{abstract}
\maketitle

\section{INTRODUCTION}

Actinide elements and compounds possess particular interesting physical
behaviors due to the 5\emph{f} states and have always attracted extensive
attention because of their importance in nuclear fuel cycle
\cite{Heathman,Atta-Fynn,Prodan2,Moore}. Actinide dioxides, \emph{A}O$_{2}$
(\emph{A}=Th, U, Np or Pu), are universally used as advanced fuel materials
for nuclear reactors and PuO$_{2}$ also plays an important role in the
plutonium re-use, separation and/or long-term storage. Recently, there has
occurred in the literature a series of experimental reports
\cite{Haschke,Butterfield,Gouder} on the strategies of storage of Pu-based
waste. Exposure to air and moisture, metallic plutonium surface easily
oxidizes to Pu$_{2}$O$_{3}$ and PuO$_{2}$. Although the existence of
PuO$_{2+x}$ ($x\leq$0.27) was reported by Haschke \emph{et al.} \cite{Haschke}%
, recent photoemission study found that PuO$_{2}$ was only covered by a
chemisorbed layer of oxygen and can be easily desorbed at elevated temperature
\cite{Gouder}, therefore, PuO$_{2}$ is the stablest plutonium oxide. At
ambient condition PuO$_{2}$ crystallizes in a fluorite structure with space
group \emph{Fm$\bar{3}$m} (No. 225). Its cubic unit cell is composed of four
PuO$_{2}$ formula units with plutonium atoms and oxygen atoms in 4\emph{a} and
8\emph{c} sites, respectively. By using the energy dispersive x-ray
diffraction method, Dancausse \emph{et al} \cite{Dancausse} reported that at
39 GPa, PuO$_{2}$ undergoes a phase transition to an orthorhombic structure of
cotunnite type with space group \emph{Pnma} (No. 62).

As for the electronic-structure study of PuO$_{2}$, conventional density
functional theory (DFT) that applies the local density approximation (LDA) or
generalized gradient approximation (GGA) underestimates the strong on-site
Coulomb repulsion of the 5\emph{f} electrons and, consequently, describes
PuO$_{2}$ as incorrect ferromagnetic (FM) conductor \cite{Boettger2} instead
of antiferromagnetic (AFM) Mott insulator reported by experiment
\cite{McNeilly}. Same problems have been confirmed in studying other
correlated materials within the pure LDA/GGA schemes. Fortunately, several
approaches, the LDA/GGA+\emph{U} \cite{Dudarev1,Dudarev2,Dudarev3}, the hybrid
density functional of HSE \cite{ProdanJCP}, and the self-interaction corrected
local spin-density (SIC-LSD) \cite{Petit}, have been developed to correct
these failures in calculations of actinide compounds. The effective
modification of pure DFT by LDA/GGA+\emph{U} formalisms has been confirmed
widely in study of UO$_{2}$ \cite{Dudarev1,Dudarev3} and PuO$_{2}$
\cite{SunJCP,SunCP,Andersson,Jomard,Shi}. By tuning the effective Hubbard
parameter in a reasonable range, the AFM Mott insulator feature was correctly
calculated and the structural parameters as well as the electronic structure
are well in accord with experiments. Lattice dynamical properties of UO$_{2}$
and PuO$_{2}$ and various contributions to their thermal conductivities were
studied by using a combination of LDA and Dynamical Mean-Field Theory (DMFT)
\cite{Yin}. However, despite that abundant researches on the structural,
electronic, optical, thermodynamic properties of plutonium dioxide have been
performed, relatively little is known regarding its chemical bonding,
mechanical properties, and phonon dispersion. In addition, the pressure
induced structural transition has also not yet been studied by
first-principles DFT+\emph{U} calculations.

In this work, we have systematically calculated the ground-state structural
parameters, electronic, mechanical, thermodynamic properties, and pressure
induced structural transition of PuO$_{2}$ by employing the LDA+\emph{U} and
GGA+\emph{U} schemes due to Dudarev \emph{et al}.
\cite{Dudarev1,Dudarev2,Dudarev3}. The validity of the ground-state
calculation is carefully tested. Our calculated lattice parameter and bulk
modulus \emph{B} are well consistent with previous LDA+\emph{U} results
\cite{Andersson}. The total energy, structural parameters, bulk modulus
\emph{B}, and pressure derivative of the bulk modulus \emph{B$^{\prime}$} for
AFM and FM phases calculated in wide range of effective Hubbard \emph{U}
parameter are presented and our electronic spectrum reproduce briefly the main
feature of our previous study \cite{SunJCP}. In addition, the bonding nature
of \emph{A}$-$O bond in PuO$_{2}$, NpO$_{2}$, UO$_{2}$, and ThO$_{2}$
involving its mixed ionic/covalent character is investigated by performing the
Bader analysis \cite{Bader,Tang}. We find that about 2.40, 2.48, 2.56, and
2.66 electrons transfer from each Pu, Np, U or Th atom to O atom,
respectively. In study of the mechanical properties, we first calculate the
elastic constants of both \emph{Fm$\bar{3}$m} and \emph{Pnma} phases and then
the elastic moduli, Poisson's ratio, and Debye temperature are deduced from
the calculated elastic constants. Hardness and ideal tensile strength of
\emph{Fm$\bar{3}$m} PuO$_{2}$ are also obtained by LDA+\emph{U} approach with
one typical value of effective \emph{U} parameter. The hardness of PuO$_{2}$
is equal to 26.6 GPa and the ideal tensile strengthes are calculated to be
81.2, 28.3, and 16.8 GPa for pulling in the [001], [110], and [111]
directions, respectively. As for the thermodynamic study, the phonon
dispersion illustrates the stability of PuO$_{2}$ and we further predict the
lattice vibration energy, thermal expansion, and specific heat by utilizing
the quasiharmonic approximation based on the first-principles phonon density
of state (DOS). One more aim of the present work is to extend the description
of PuO$_{2}$ to high pressures. The electronic, structural, and magnetic
behavior of PuO$_{2}$ under pressure up to 280 GPa have been evaluated.
Results show that there occurs a metallic transition at around 133 GPa for
\emph{Pnma} phase. The isostructural transition, similar to UO$_{2}$
\cite{Geng} and ThO$_{2}$ \cite{WangThO2}, in pressure domain of 75-133 GPa is
predicted. The rest of this paper is arranged as follows. In Sec. II the
computational method is briefly described. In Sec. III we present and discuss
our results. In Sec. IV we summarize the conclusions of this work.

\section{computational methods}

The DFT calculations are performed on the basis of the frozen-core projected
augmented wave (PAW) method of Bl\"{o}chl \cite{PAW} encoded in Vienna
\textit{ab initio} simulation package (VASP) \cite{Kresse3} using the LDA and
GGA \cite{LDA,GGA}. For the plane-wave set, a cutoff energy of 500 eV is used.
The \emph{k}-point meshes in the full wedge of the Brillouin zone (BZ) are
sampled by 9$\times$9$\times$9 and 9$\times$15$\times$9 grids according to the
Monkhorst-Pack \cite{Monk} scheme for fluorite and cotunnite PuO$_{2}$,
respectively, and all atoms are fully relaxed until the Hellmann-Feynman
forces become less than 0.02 eV/\AA . The plutonium 6\emph{s}$^{2}$%
7\emph{s}$^{2}$6\emph{p}$^{6}$6\emph{d}$^{2}$5\emph{f}$^{4}$ and the oxygen
2\emph{s}$^{2}$2\emph{p}$^{4}$ electrons are treated as valence electrons. The
strong on-site Coulomb repulsion among the localized Pu 5\emph{f} electrons is
described by using the LDA/GGA+\emph{U} formalisms formulated by Dudarev
\emph{et al.} \cite{Dudarev1,Dudarev2,Dudarev3}, where the double counting
correction has already been included. As concluded in some previous studies of
actinide dioxides (\emph{A}O$_{2}$), although the pure LDA and GGA fail to
depict the electronic structure, especially the insulating nature and the
occupied-state character of UO$_{2}$ \cite{DudarevUO2,Geng}, NpO$_{2}$
\cite{Andersson,WangNpO2}, and PuO$_{2}$ \cite{SunJCP,SunCP,Jomard}, the
LDA/GGA+\emph{U} approaches will capture the Mott insulating properties of the
strongly correlated U 5\emph{f}, Np 5\emph{f}, and Pu 5\emph{f} electrons in
AO$_{2}$ adequately. In this paper the Coulomb $U$ is treated as a variable,
while the exchange energy is set to be a constant $J$=0.75 eV. This value of
$J$ is same with our previous study of plutonium oxides \cite{SunJCP,SunCP}.
Since only the difference between $U$ and $J$ is significant \cite{Dudarev2},
thus we will henceforth label them as one single parameter, for simplicity
labeled as $U$, while keeping in mind that the non-zero $J$ has been used
during calculations.

\begin{figure}[ptb]
\begin{center}
\includegraphics[width=0.5\linewidth]{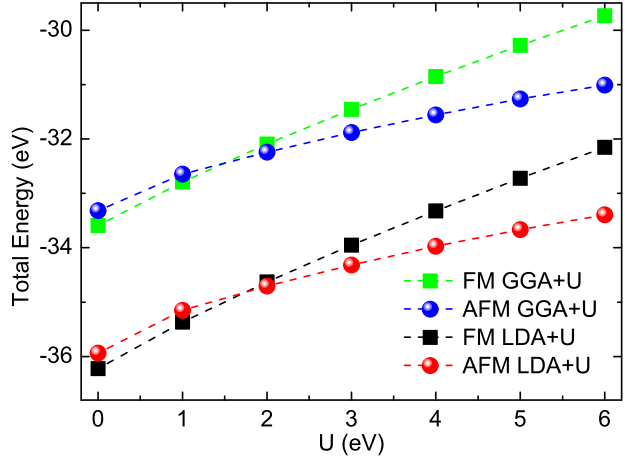}
\end{center}
\caption{(Color online) Dependence of the total energies (per formula unit) on
\emph{U} for FM and AFM PuO$_{2}$.}%
\label{toten}%
\end{figure}

Both spin-unpolarized and spin-polarized calculations are performed in this
study. Compared to FM and AFM phases, the nonmagnetic (NM) phase is not
energetically favorable both in the LDA+\emph{U} and GGA+\emph{U} formalisms.
Therefore, the results of NM are not presented in the following. The
dependence of the total energy (per formula unit at respective optimum
geometries) on \emph{U} for both FM and AFM phases within the LDA+\emph{U} and
GGA+\emph{U} formalisms is shown in Fig. \ref{toten}. At \emph{U}=0 and 1.0
eV, the total energy of the FM phase is lower than that of the AFM phase
either in LDA+\emph{U} scheme or GGA+\emph{U} scheme. However, as shown in
Fig. 1, it is clear that the total energy of the AFM phase decreases to become
lower than that of the FM phase when increasing \emph{U}. The total-energy
differences ($E_{\text{FM}}\mathtt{-}E_{\text{AFM}}$) within the LDA+\emph{U}
and GGA+\emph{U} at \emph{U}=4 eV are 0.705 and 0.651 eV, respectively. Both
FM and AFM results will be presented in the following analysis. Besides, while
the spin-orbit coupling (SOC) is important for certain properties of heavy
metal compounds, it has been numerically found \cite{Boettger1,Boettger2} and
physically analyzed \cite{ProdanJCP,Prodan1} that the inclusion of the SOC has
little effect on the bulk and one-electron properties of UO$_{2}$ and
PuO$_{2}$. Our test calculations also show that within LDA+\emph{U} approach
with \emph{U}=4 eV, inclusion of SOC will increase the optimum lattice
constant by only 0.7\% and the bulk modulus by about 0.5 GPa. Therefore, in
our following calculations of plutonium dioxide, the SOC is not included.

In present work, the theoretical equilibrium volume, bulk modulus \emph{B},
and pressure derivative of the bulk modulus \emph{B$^{\prime}$} are obtained
by fitting the energy-volume data with the third-order Birch-Murnaghan
equation of state (EOS) \cite{Birch}. In order to calculate elastic constants,
a small strain is applied onto the structure. For small strain $\epsilon$,
Hooke's law is valid and the crystal energy $E(V,\epsilon)$ can be expanded as
a Taylor series \cite{Nye},
\begin{align}
E(V,\epsilon)=E(V_{0},0)+V_{0}\sum_{i=1}^{6}\sigma_{i}e_{i}+\frac{V_{0}}%
{2}\sum_{i,j=1}^{6}C_{ij}e_{i}e_{j}+O(\{e_{i}^{3}\}),
\end{align}
where $E(V_{0},0)$ is the energy of the unstrained system with the equilibrium
volume $V_{0}$, $\epsilon$ is strain tensor which has matrix elements
$\varepsilon_{ij}$ ($i,j$=1, 2, and 3) defined by
\begin{align}
\varepsilon_{ij}=\left(
\begin{array}
[c]{ccc}%
e_{1} & \frac{1}{2}e_{6} & \frac{1}{2}e_{5}\\
\frac{1}{2}e_{6} & e_{2} & \frac{1}{2}e_{4}\\
\frac{1}{2}e_{5} & \frac{1}{2}e_{4} & e_{3}%
\end{array}
\right)  ,
\end{align}
and $C_{ij}$ are the elastic constants. For cubic structures, there are three
independent elastic constants ($C_{11}$, $C_{12}$, and $C_{44}$). So, the
elastic constants for fluorite PuO$_{2}$ can be calculated from three
different strains listed as follows:
\begin{align}
\emph{$\epsilon$$^1$}=(\delta,\delta,\delta,0,0,0), \emph{$\epsilon$$%
^2$}=(\delta,0,\delta,0,0,0), \emph{$\epsilon$$^3$}=(0,0,0,\delta
,\delta,\delta).
\end{align}
The strain amplitude $\delta$ is varied in steps of 0.006 from $\delta$%
=$-$0.036 to 0.036 and the total energies $E(V,\delta)$ at these strain steps
are calculated. After obtaining elastic constants, we can calculate bulk and
shear moduli from the Voigt-Reuss-Hill (VRH) approximations \cite{Hill}. The
Voigt (Reuss) bounds on the bulk modulus \emph{B$_{V}$} (\emph{B$_{R}$}) and
shear modulus \emph{G$_{V}$} (\emph{G$_{R}$}) for this cubic crystal system
are deduced from the formulae of elastic moduli in Ref. \cite{Hanies}. As for
cotunnite PuO$_{2}$, the nine independent elastic constants ($C_{11}$,
$C_{12}$, $C_{13}$, $C_{22}$, $C_{23}$, $C_{33}$, $C_{44}$, $C_{55}$, and
$C_{66}$) can be obtained from nine different strains listed in the
following:
\begin{align}
&  \emph{$\epsilon$$^1$}=(\delta,0,0,0,0,0), \emph{$\epsilon$$^2$}%
=(0,\delta,0,0,0,0), \emph{$\epsilon$$^3$}=(0,0,\delta,0,0,0),\nonumber\\
&  \emph{$\epsilon$$^4$}=(0,0,0,\delta,0,0), \emph{$\epsilon$$^5$}%
=(0,0,0,0,\delta,0), \emph{$\epsilon$$^6$}=(0,0,0,0,0,\delta),\nonumber\\
&  \emph{$\epsilon$$^7$}=(\delta,\delta,0,0,0,0), \emph{$\epsilon$$%
^8$}=(0,\delta,\delta,0,0,0), \emph{$\epsilon$$^9$}=(\delta,0,\delta,0,0,0)
\end{align}
and the formulae of elastic moduli in VRH approximations \cite{Hill} are from
Ref. \cite{Watt}. Based on the Hill approximation \cite{Hill}, \emph{B}%
=$\frac{1}{2}(B_{R}+B_{V})$ and \emph{G}=$\frac{1}{2}(G_{R}+G_{V})$. The
Young's modulus \emph{E} and Poisson's ratio $\upsilon$ are given by the
following formulae:
\begin{align}
E=9BG/(3B+G), \upsilon=(3B-2G)/[2(3B+G)].
\end{align}

In addition, the elastic properties of a solid can also be related to
thermodynamical parameters especially specific heat, thermal expansion, Debye
temperature, melting point, and Gr\"{u}neisen parameter \cite{Ravindran}. From
this point of view, Debye temperature is one of fundamental parameters for
solid materials. Due to the fact that at low temperatures the vibrational
excitations arise solely from acoustic vibrations, therefore, the Debye
temperature calculated from elastic constants is the same as that determined
from specific heat measurements. The relation between the Debye temperature
($\theta_{D}$) and the average sound wave velocity ($\upsilon_{m}$) is
\begin{equation}
\theta_{D}=\frac{h}{k_{B}}\left(  \frac{3n}{4\pi\Omega}\right)  ^{1/3}%
\upsilon_{m},
\end{equation}
where \emph{h} and $\emph{k}_{B}$ are Planck and Boltzmann constants,
respectively, \emph{n} is the number of atoms in the molecule and $\Omega$ is
molecular volume. The average wave velocity in the polycrystalline materials
is approximately given as
\begin{equation}
\upsilon_{m}=\left[  \frac{1}{3}\left(  \frac{2}{\upsilon_{t}^{3}}+\frac
{1}{\upsilon_{l}^{3}}\right)  \right]  ^{-1/3},
\end{equation}
where $\upsilon_{t}$=$\sqrt{G/\rho}$ ($\rho$ is the density) and $\upsilon
_{l}$=$\sqrt{(3B+4G)/3\rho}$ are the transverse and longitudinal elastic wave
velocities of the polycrystalline materials, respectively.

\section{results}

\begin{table}[ptb]
\caption{Calculated lattice parameters (\emph{a}$_{0}$), bulk modulus
(\emph{B}), and pressure derivative of the bulk modulus (\emph{B$^{^{\prime}}%
$}) for AFM and FM PuO$_{2}$ at 0 GPa. For comparison, experimental values are
also listed.}%
\label{latticeTable}
\begin{ruledtabular}
\begin{tabular}{cccccccccccccccc}
Magnetism&Method&property&\emph{U}=0&\emph{U}=1&\emph{U}=2&\emph{U}=3&\emph{U}=4&\emph{U}=5&\emph{U}=6&Expt.\\
\hline
AFM&LDA+\emph{U}&\emph{a}$_{0}$ ({\AA})&5.275&5.313&5.338&5.351&5.362&5.371&5.378&5.398$^{\emph{a}}$\\
&&\emph{B} (GPa)&218&208&224&224&225&226&227&178$^{\emph{b}}$\\
&&\emph{B$^{^{\prime}}$}&4.1&3.7&4.3&4.3&4.3&4.3&4.3&\\
&GGA+\emph{U}&\emph{a}$_{0}$ ({\AA})&5.396&5.433&5.446&5.457&5.466&5.473&5.480&\\
&&\emph{B} (GPa)&185&188&191&192&193&194&195&\\
&&\emph{B$^{^{\prime}}$}&4.3&3.8&4.4&4.5&4.5&4.5&4.4&\\
FM&LDA+\emph{U}&\emph{a}$_{0}$ ({\AA})&5.270&5.290&5.309&5.325&5.338&5.350&5.361&\\
&&\emph{B} (GPa)&230&224&221&220&218&215&212&\\
&&\emph{B$^{^{\prime}}$}&4.4&4.4&4.3&4.4&4.4&4.4&4.5&\\
&GGA+\emph{U}&\emph{a}$_{0}$ ({\AA})&5.384&5.405&5.424&5.439&5.452&5.464&5.476&\\
&&\emph{B} (GPa)&193&188&186&184&182&179&174&\\
&&\emph{B$^{^{\prime}}$}&4.5&4.4&4.5&4.5&4.7&4.7&4.8&\\
\end{tabular}
$^{\emph{a}}$ Reference \cite{Haschke}, $^{\emph{b}}$ Reference
\cite{Idiri}.
\end{ruledtabular}
\end{table}

\subsection{Atomic and electronic structures of fluorite PuO$_{2}$}

We report in Table \ref{latticeTable} the lattice parameter (\emph{a}$_{0}$),
bulk modulus (\emph{B}), and pressure derivative of the bulk modulus
(\emph{B$^{^{\prime}}$}) for AFM and FM PuO$_{2}$ obtained in LDA+\emph{U} and
GGA+\emph{U} frameworks. All these values are determined by EOS fitting. For
comparison, the experimental values of $a_{0}$ (Ref. \cite{Haschke}) and
\emph{B} (Ref. \cite{Idiri}) are also listed. In the overall view, the
dependence of the lattice parameter $a_{0}$ of AFM PuO$_{2}$ on \emph{U} is
well consistent with our previous study \cite{SunJCP} either in LDA+\emph{U}
scheme or in GGA+\emph{U} scheme. For the LDA/GGA+\emph{U} approaches, the
calculated $a_{0}$ improves upon the pure LDA/GGA by steadily increasing its
amplitude with \emph{U}. Actually, at a typical value \emph{U}=4 eV, the
LDA+\emph{U} gives $a_{0}$=5.362 \AA , which is very close to the experimental
value of 5.398 \AA , and the GGA+\emph{U} gives $a_{0}$=5.466 \AA . Note that
recent PBE+\emph{U} \cite{Jomard} and LDA+\emph{U} \cite{Andersson}
calculations with \emph{U}=4.0 eV and \emph{J}=0.7 eV predicted the lattice
parameter of AFM PuO$_{2}$ to be 5.444 and 5.354 \AA , respectively, and the
HSE \cite{ProdanJCP} and SIC-LSD \cite{Petit} calculations gave the values to
be 5.385 and 5.44 \AA , respectively. On the other hand, as shown in Table
\ref{latticeTable}, the tendency of $a_{0}$ with \emph{U} for FM PuO$_{2}$ is
similar to that of the AFM phase. At a typical value \emph{U}=4 eV, the
LDA/GGA+\emph{U} give $a_{0}$=5.338 and 5.452 \AA . These values are slightly
smaller than those of the AFM phase. Previous HSE \cite{ProdanJCP} calculation
of FM PuO$_{2}$ gave the lattice parameter to be 5.387 \AA . As for the
dependence of bulk modulus \emph{B} of AFM and FM PuO$_{2}$ on \emph{U}, it is
clear that the LDA results are always higher than the GGA results, which is
due to the overbinding effect of the LDA approach. While the bulk modulus
\emph{B} of AFM phase increases steadily with increasing the amplitude of
\emph{U}, the bulk modulus \emph{B} of FM phase will decrease along with the
increasing of Hubbard \emph{U}. At a typical value \emph{U}=4 eV, the
LDA+\emph{U} and GGA+\emph{U} give \emph{B}=225 and 193 GPa for AFM phase,
respectively, and \emph{B}=218 and 182 GPa for FM phase, respectively.
Apparently, the GGA+\emph{U} approach gives more close values to the
experimental data of \emph{B}=178 GPa \cite{Idiri}. In our present study, as
shown in Table \ref{elastic}, the bulk modulus \emph{B} deduced from elastic
constants turns out to be very close to that obtained by EOS fitting. This
indicates that our calculations are consistent and reliable. In addition, as
listed in Table \ref{elastic}, recent PBE+\emph{U} \cite{Jomard} and
LDA+\emph{U} \cite{Andersson} calculations with \emph{U}=4.0 eV and
\emph{J}=0.7 eV predicted the bulk modulus \emph{B} of AFM PuO$_{2}$ to be 199
and 226 GPa, respectively, and the HSE \cite{ProdanJCP} and SIC-LSD
\cite{Petit} calculations gave the the bulk modulus to be 221 and 214 GPa,
respectively. As for pressure derivative of the bulk modulus
(\emph{B$^{^{\prime}}$}), all our calculated values are approximately equal to
4.5. Overall, comparing with the experimental data and recent theoretical
results, the accuracy of our atomic-structure prediction for AFM PuO$_{2}$ is
quite satisfactory by tuning the effective Hubbard parameter \emph{U} in a
range of 3-4 eV within the LDA/GGA+\emph{U} approaches, which supplies the
safeguard for our following study of electronic structure and mechanical
properties of PuO$_{2}$.

\begin{table}[ptb]
\caption{Calculated elastic constants, various moduli, Poisson's ratio
($\upsilon$), spin moments ($\mu_{mag.}$), and energy band gap (\emph{E$_{g}$%
}) for \emph{Fm$\bar{3}$m} AFM PuO$_{2}$ at 0 GPa. For comparison,
experimental values and other theoretical results are also listed.}%
\label{elastic}
\begin{ruledtabular}
\begin{tabular}{cccccccccccccccc}
Method&\emph{C$_{11}$}&\emph{C$_{12}$}&\emph{C$_{44}$}&\emph{B}&\emph{G}&\emph{E}&$\upsilon$&$\mu_{mag.}$&E$_{g}$\\
&(GPa)&(GPa)&(GPa)&(GPa)&(GPa)&(GPa)&&($\mu_{B}$)&(eV)\\
\hline
LDA+\emph{U} (\emph{U}=0)&386.6&136.5&71.9&220&89.9&237.3&0.320&3.937&0.0\\
LDA+\emph{U} (\emph{U}=4)&319.6&177.8&74.5&225&73.0&197.7&0.354&4.126&1.5\\
GGA+\emph{U} (\emph{U}=0)&343.9&112.3&53.7&190&73.5&195.1&0.328&4.085&0.0\\
GGA+\emph{U} (\emph{U}=4)&256.5&167.9&59.2&197&52.7&145.2&0.377&4.165&1.5\\
Expt.& &  &  &178$^{\emph{a}}$&&&&&1.8$^{\emph{b}}$\\
PBE+\emph{U}$^{\emph{c}}$& &  &  &199&&&&3.89&2.2\\
LDA+\emph{U}$^{\emph{d}}$& &  &  &226&&&&&1.7\\
HSE$^{\emph{e}}$& &  &  &221&&&&&3.4\\
SIC-LSD$^{\emph{f}}$& &  &  &214&&&&&1.2\\
\end{tabular}
$^{\emph{a}}$ Reference \cite{Idiri}, $^{\emph{b}}$ Reference
\cite{McNeilly}, $^{\emph{c}}$ Reference \cite{Jomard},
$^{\emph{d}}$ Reference \cite{Andersson},$^{\emph{e}}$ Reference
\cite{ProdanJCP}, $^{\emph{f}}$ Reference \cite{Petit}.
\end{ruledtabular}
\end{table}

\begin{figure}[ptb]
\begin{center}
\includegraphics[width=0.8\linewidth]{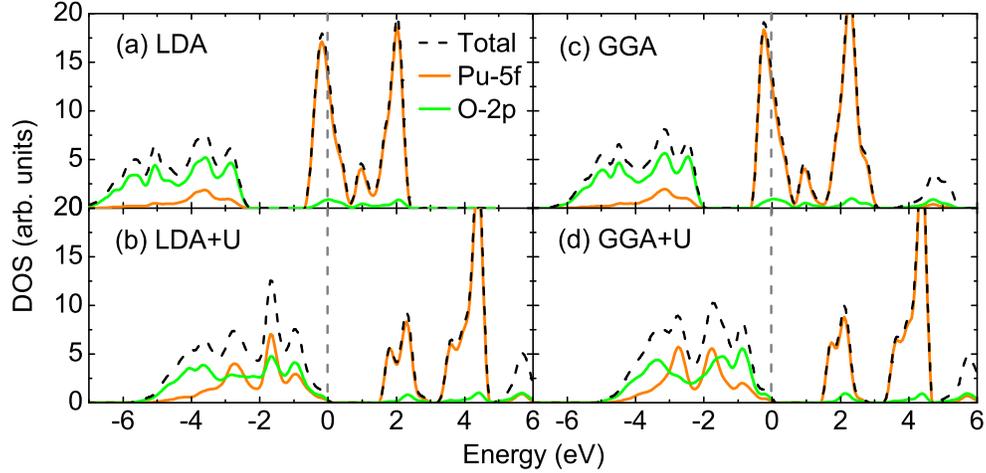}
\end{center}
\caption{(Color online) The total DOS for the PuO$_{2}$ AFM phase computed in
the (a) LDA, (b) LDA+\emph{U} (\emph{U}=4), (c) GGA, and (d) GGA+\emph{U}
(\emph{U}=4) formalisms. The projected DOSs for the Pu 5\emph{f} and O
2\emph{p} orbitals are also shown. The Fermi energy level is set at zero.}%
\label{DOS}%
\end{figure}

\begin{figure}[ptb]
\begin{center}
\includegraphics[width=0.5\linewidth]{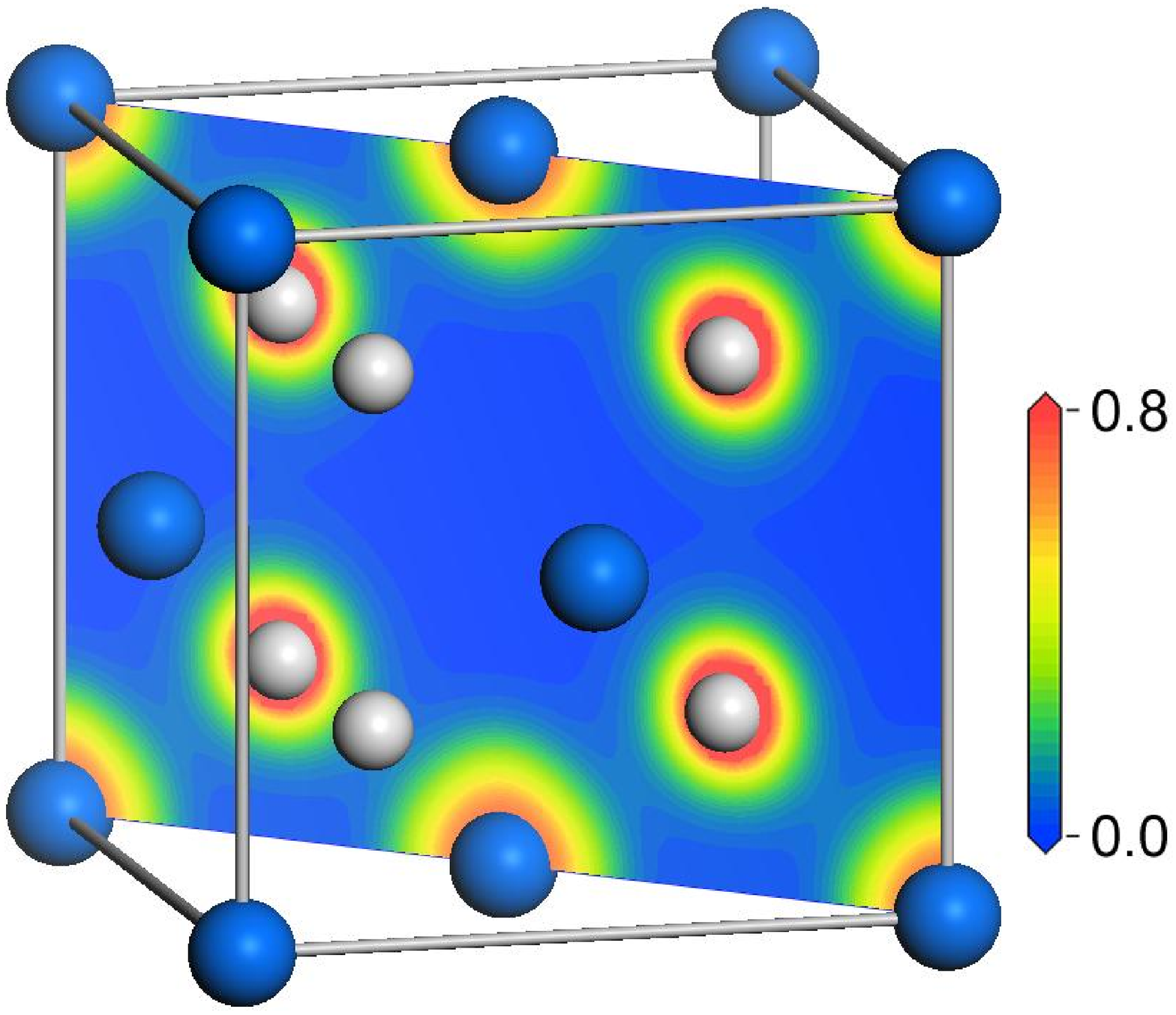}
\end{center}
\caption{(Color online) Cubic unit cell for PuO$_{2}$ in space group
\emph{Fm$\bar{3}$m}. The larger blue spheres represent plutonium
atoms and the smaller white O atoms. A slice of the isosurfaces of
the electron density for AFM phase in (110) plane calculated in the
LDA+\emph{U} formalisms with \emph{U}=4 eV is also presented (in
unit of e/{\AA }$^{3}$)}%
\label{charge}%
\end{figure}

The total density of states (DOS) as well as the projected DOS for the Pu
5\emph{f} and O 2\emph{p} orbitals within LDA, GGA, LDA+\emph{U}, and
GGA+\emph{U} formalisms are shown in Fig. \ref{DOS}. Clearly, our results
reproduce all the features included in our previous work \cite{SunJCP}. In
particular, we recover the main conclusion that although the pure LDA and GGA
fail to depict the electronic structure, especially the insulating nature and
the occupied-state character of PuO$_{2}$, by tuning the effective Hubbard
parameter in a reasonable range, the LDA/GGA+\emph{U} approaches can
prominently improve upon the pure LDA/GGA calculations and, thus, can provide
a satisfactory qualitative electronic structure description comparable with
the photoemission experiments \cite{Butterfield,Gouder}. In this study, the
insulating energy band gap (\emph{E}$_{g}$) is of 1.5 eV at \emph{U}=4 eV
within LDA/GGA+\emph{U} approaches (see Table \ref{elastic}). In Table
\ref{elastic}, the experimental \cite{McNeilly} value and previous theoretical
calculations \cite{Jomard,Andersson,ProdanJCP,Petit} results are also listed
for comparison. Obviously, the HSE calculations result in a larger
\emph{E}$_{g}$ by $\mathtt{\sim}$1.9 eV \cite{ProdanJCP} and our calculated
results are well consistent with the measured value and other theoretical
results. The calculated amplitude of local spin moment is $\sim$4.1 $\mu_{B}$
(per Pu atom) for AFM PuO$_{2}$ within the two DFT+\emph{U} schemes and this
value is comparable to previous LDA/PBE+\emph{U} \cite{Jomard} results of
$\sim$3.9 $\mu_{B}$.

\begin{table}[ptb]
\caption{Calculated charge and volumes according to Bader partitioning as well
as the \emph{A}$-$O distances and correlated minimum values of charge density
along the \emph{A}$-$O bonds for actinide dioxides.}%
\label{bader}
\begin{ruledtabular}
\begin{tabular}{cccccccccccccc}
Compound&Q$_{B}$(\emph{A})&Q$_{B}$(O)&V$_{B}$(\emph{A})&V$_{B}$(O)&\emph{A}$-$O&Charge density$_{\rm{min.}}$\\
&($e$)&($e$)&({\AA}$^{3}$)&({\AA}$^{3}$)&({\AA})&($e$/{\AA}$^{3}$)\\
\hline
PuO$_{2}$&13.60&7.20&15.36&11.58&2.32&0.53\\
NpO$_{2}$&12.52&7.24&15.69&11.83&2.34$^{\emph{a}}$&0.51$^{\emph{a}}$\\
UO$_{2}$&11.44&7.28&16.35&12.05&2.36&0.52\\
ThO$_{2}$&9.34&7.33&17.67&13.35&2.43$^{\emph{b}}$&0.45$^{\emph{b}}$\\
\end{tabular}
\end{ruledtabular}
$^{\emph{a}}$ Reference \cite{WangNpO2}, $^{\emph{b}}$ Reference
\cite{WangThO2}.\end{table}

To understand the chemical bonding characters of fluorite PuO$_{2}$, we
present in Fig. \ref{charge} the crystal structure of its cubic unit cell and
the charge density map of the (110) plane calculated in LDA+\emph{U} formalism
with \emph{U}=4 eV for AFM phase. Evidently, the charge density around Pu and
O ions are all near spherical distribution with slightly deformed toward the
direction to their nearest neighboring atoms and there are clear covalent
bridges between Pu and O ions. In order to describe the ionic/covalent
character quantitatively and more clearly, we plot the line charge density
distribution along the nearest Pu$-$O bonds (not shown). A minimum value of
charge density (0.53 $e$/{\AA }$^{3}$) along the Pu$-$O bonds is obtained and
is listed in Table \ref{bader}. For comparison, we have also calculated some
corresponding properties of AFM UO$_{2}$ within the LDA+\emph{U} formalism.
Parameters of the Hubbard term are taken as \emph{U}=4.51 eV and \emph{J}=0.51
eV, which had been checked carefully by Dudarev \emph{et al}
\cite{Dudarev1,Dudarev2,Dudarev3}. In following study, results of UO$_{2}$ are
all calculated using these Hubbard parameters either for ground-state
\emph{Fm$\bar{3}$m} phase or high pressure \emph{Pnma} phase. The lattice
parameter $a_{0}$=5.449 \AA \ and bulk modulus \emph{B}=220.0 GPa for
\emph{Fm$\bar{3}$m} UO$_{2}$ obtained by EOS fitting are in perfect agreement
with results of recent LDA+\emph{U} calculation \cite{Andersson} ($a_{0}%
$=5.448 \AA \ and \emph{B}=218 GPa) and experiments \cite{Yamashita,Idiri}
($a_{0}$=5.47 \AA \ and \emph{B}=207 GPa). Charge analysis results of
\emph{Fm$\bar{3}$m} UO$_{2}$ are listed in Table \ref{bader}. Clearly, the
minimum values of charge density for PuO$_{2}$ and UO$_{2}$, comparable to
that along the Np$-$O bonds included in our previous study of NpO$_{2}$
\cite{WangNpO2}, are prominently larger than that along the Th$-$O bonds (0.45
e/{\AA }$^{3}$) in ThO$_{2}$ \cite{WangThO2}. This indicates that the Pu$-$O,
U$-$O, and Np$-$O bonds have stronger covalency than the Th$-$O bonds. And
this conclusion is in good accordance with previous HSE study of the covalency
in Ref. \cite{Prodan2}, in which significant orbital mixing and covalency in
the intermediate region (PuO$_{2}$-CmO$_{2}$) of actinide dioxides was for the
first time showed by the increasing 5$f$-2$p$ orbital energy degeneracy.
Besides, the \emph{A}$-$O (\emph{A}=Pu, Np, U or Th) bond distances calculated
in present study and our previous works \cite{WangNpO2,WangThO2} are listed in
Table \ref{bader}. Obviously, the Pu$-$O, Np$-$O, and U$-$O bond distances are
all smaller than the Th$-$O bond distance. This fact illustrates that the
covalency of \emph{A}O$_{2}$ has tight relation with their bond distances,
thus influences their macroscopical properties, such as hardness and
elasticity. To see the ionicity of \emph{A}O$_{2}$, results from the Bader
analysis \cite{Bader,Tang} are presented in Table \ref{bader}. The charge
(Q$_{B}$) enclosed within the Bader volume (V$_{B}$) is a good approximation
to the total electronic charge of an atom. Note that although we have included
the core charge in charge density calculations, since we do not expect
variations as far as the trends are concerned, only the valence charge are
listed. Here, same with our previous study, we apply \emph{U}=4.6 eV and
\emph{J}=0.6 eV for the Np 5\emph{f} orbitals in calculations of NpO$_{2}$.
Based on the data in Table \ref{bader} and considering the valence electron
numbers of Pu, Np, U, Th, and O atoms (16, 15, 14, 12, and 6, respectively),
in our study of \emph{A}O$_{2}$, we find that about 2.40, 2.48, 2.56, and 2.66
electrons transfer from each Pu, Np, U or Th atom to O atom, respectively.
This indicates that the ionicity of AO$_{2}$ shows decreasing trend with
increasing \emph{Z}.

\subsection{Mechanical properties}

\subsubsection{Elastic properties}

Our calculated elastic constants, various moduli, and Poisson's ratio
$\upsilon$ of the \emph{Fm$\bar{3}$m} PuO$_{2}$ at 0 GPa are collected in
Table \ref{elastic} and those of the \emph{Pnma} phase are listed in Table
\ref{Pnmaelastic}. Obviously, the \emph{Fm$\bar{3}$m} phase of PuO$_{2}$ is
mechanically stable due to the fact that its elastic constants satisfy the
following mechanical stability criteria \cite{Nye} of cubic structure:
\begin{align}
C_{11}>0, C_{44}>0, C_{11}>|C_{12}|, (C_{11}+2C_{12})>0.
\end{align}
As for the high-pressure \emph{Pnma} crystalline phase, we have optimized the
structural parameters of its AFM phase at different pressures within
LDA+\emph{U} formalism with the typical value of \emph{U}=4 eV. To avoid the
Pulay stress problem, we perform the structure relaxation calculations at
fixed volumes rather than constant pressures. Note that different from the
structure optimization of the ground-state \emph{Fm$\bar{3}$m} phase, the
coordinates of atoms and the cell shape of the \emph{Pnma} phase are necessary
to be optimized due to its internal degrees of freedom. After fitting the
energy-volume data to the EOS, we obtain the optimized structural lattice
parameters \emph{a}, \emph{b}, and \emph{c} for the \emph{Pnma} PuO$_{2}$ AFM
phase at 0 GPa to be 5.889, 3.562, and 6.821 {\AA }, respectively, giving
\emph{V}=143.1 {\AA }$^{3}$. This volume is smaller than the equilibrium
volume of 154.2 {\AA }$^{3}$ for the \emph{Fm$\bar{3}$m} phase. This feature
implies that the \emph{Pnma} phase will become stable under compression. The
elastic constants listed in Table \ref{Pnmaelastic} indicate that the
\emph{Pnma} PuO$_{2}$ is also mechanically stable. Actually, they satisfy the
following mechanical stability criteria \cite{Nye} for the orthorhombic
structure:
\begin{align}
&  C_{11}>0,C_{22}>0,C_{33}>0,C_{44}>0,C_{55}>0,C_{66}>0,\nonumber\\
&  [C_{11}+C_{22}+C_{33}+2(C_{12}+C_{13}+C_{23})]>0,\nonumber\\
&  (C_{11}+C_{22}-2C_{12})>0,(C_{11}+C_{33}-2C_{13})>0,\nonumber\\
&  (C_{22}+C_{33}-2C_{23})>0.
\end{align}
\begin{table}[ptb]
\caption{Calculated elastic constants, elastic moduli, pressure derivative of
the bulk modulus \emph{B$^{^{\prime}}$}, and Poisson's ratio $\upsilon$ for
cotunnite-type PuO$_{2}$ and UO$_{2}$ at 0 GPa. Except \emph{B$^{^{\prime}}$}
and $\upsilon$, all other values are in units of GPa.}%
\label{Pnmaelastic}
\begin{ruledtabular}
\begin{tabular}{lccccccccccccccccc}
Compound&\emph{C$_{11}$}&\emph{C$_{22}$}&\emph{C$_{33}$}&\emph{C$_{44}$}&\emph{C$_{55}$}&\emph{C$_{66}$}&\emph{C$_{12}$}&\emph{C$_{23}$}&\emph{C$_{13}$}&\emph{B}&\emph{B$^{'}$}&\emph{B$_{a}$}&\emph{B$_{b}$}&\emph{B$_{c}$}&\emph{G}&\emph{E}&$\upsilon$\\
\hline
PuO$_{2}$&355.6 & 327.4  & 336.3 & 36.0  & 81.5 &96.7&140.5&199.0&141.2&219.9&5.7&598.5&669.5&720.3&73.0&197.2&0.351\\
UO$_{2}$&338.2 & 335.1  & 325.9 & 24.4  & 75.6 &92.0&119.9&148.4&142.1&202.2&5.5&586.4&593.8&641.9&68.0&183.4&0.349\\
\end{tabular}
\end{ruledtabular}
\end{table}One can see from Table \ref{Pnmaelastic} that the calculated
$C_{12}$, $C_{23}$, and $C_{13}$ are largely smaller than $C_{11}$, $C_{22}$,
and $C_{33}$. Therefore, the mechanical stability criteria is easily
satisfied. For comparison, we have also calculated the elastic properties of
UO$_{2}$ in its ground-state fluorite phase and high-pressure cotunnite phase
within LDA+\emph{U} formalism. For \emph{Fm$\bar{3}$m} phase, \emph{C$_{11}$%
}=389.3 GPa, \emph{C$_{12}$}=138.9 GPa, \emph{C$_{44}$}=71.3 GPa,
\emph{B}=222.4 GPa, \emph{G}=89.5 GPa, \emph{E}=236.8 GPa, and Poisson's ratio
$\upsilon$=0.323. Results for \emph{Pnma} UO$_{2}$ are presented in Table
\ref{Pnmaelastic}. Clearly, both fluorite and cotunnite phases of UO$_{2}$ are
mechanically stable. Moreover, comparing results of bulk modulus \emph{B},
shear modulus \emph{G}, Young's modulus \emph{E}, and Poisson's ratio
$\upsilon$ for fluorite phase and cotunnite phase, they are almost equal to
each other for PuO$_{2}$. Particularly, the bulk modulus \emph{B} is only
smaller by about 5 GPa for \emph{Pnma} phase compared to that of
\emph{Fm$\bar{3}$m} phase. As for UO$_{2}$, bulk modulus of cotunnite phase is
smaller by approximately 9\%, shear modulus and Young's modulus about 23\%,
compared to those of fluorite structure. Besides, in our previous study of
ThO$_{2}$ \cite{WangThO2} we find that the bulk modulus, shear modulus, and
Young's modulus of cotunnite ThO$_{2}$ are all smaller by approximately 25\%
compared to those of fluorite ThO$_{2}$. Therefore, after comparing the bulk
modulus \emph{B} for the two phases of ThO$_{2}$, UO$_{2}$, and PuO$_{2}$, we
find that the difference is decreasing along with increasing of 5$f$
electrons. This trend is understandable. Although the softening in bulk
modulus upon phase transition for these three systems is similar, the increase
of 5$f$ electrons of actinide metal atoms will lead to more covalency and thus
reduce the bulk modulus difference between the two phases of the actinide
dioxides across the series. To study the anisotropy of the linear bulk modulus
for \emph{Pnma} PuO$_{2}$ and UO$_{2}$, we calculate the directional bulk
modulus along the \textbf{a}, \textbf{b} and \textbf{c} axis by the following
equations \cite{Ravindran},
\begin{align}
\emph{B}_{a}=\frac{\Lambda}{1+\alpha+\beta}, \emph{B}_{b}=\frac{\emph{B}_{a}%
}{\alpha}, \emph{B}_{c}=\frac{\emph{B}_{a}}{\beta},
\end{align}
where $\Lambda$=$C_{11}$+2$C_{12}$$\alpha$+$C_{22}$$\alpha$$^{2}$+2$C_{13}%
$$\beta$+$C_{33}$$\beta$$^{2}$+2$C_{23}$$\alpha$$\beta$. For orthorhombic
crystals,
\begin{align}
\alpha=\frac{(C_{11}-C_{12})(C_{33}-C_{13})-(C_{23}-C_{13})(C_{11}-C_{13}%
)}{(C_{33}-C_{13})(C_{22}-C_{12})-(C_{13}-C_{23})(C_{12}-C_{23})}%
\end{align}
and
\begin{align}
\beta=\frac{(C_{22}-C_{12})(C_{11}-C_{13})-(C_{11}-C_{12})(C_{23}-C_{12}%
)}{(C_{22}-C_{12})(C_{33}-C_{12})-(C_{12}-C_{23})(C_{13}-C_{23})}%
\end{align}
are defined as the relative change of the \textbf{b} and \textbf{c} axis as a
function of the deformation of the \textbf{a} axis. After calculation, we
obtain values of $\alpha$ and $\beta$ to be 0.8939 (0.9875) and 0.8309
(0.9135) for PuO$_{2}$ (UO$_{2}$), respectively. Results of $\emph{B}_{a}$,
$\emph{B}_{b}$, and $\emph{B}_{c}$ are presented in Table \ref{Pnmaelastic}.
Clearly, results of the directional bulk moduli show that both \emph{Pnma}
PuO$_{2}$ and UO$_{2}$ are easily compressed along \textbf{a} axis at 0 GPa.
The longest axis \textbf{c} is the hardest axis for these two actinide
dioxides. Directional bulk moduli of PuO$_{2}$ are all bigger than the
corresponding values of UO$_{2}$. Moreover, using results of elastic constants
included in previous study \cite{WangThO2}, the directional bulk moduli
$\emph{B}_{a}$, $\emph{B}_{b}$, and $\emph{B}_{c}$ of \emph{Pnma} ThO$_{2}$
are calculated to be 528.3, 406.3, and 415.8 GPa, respectively. This
illustrates that, in contrary to \emph{Pnma} PuO$_{2}$ and UO$_{2}$, the
\emph{Pnma} ThO$_{2}$ is relatively harder to be compressed along \textbf{a}
axis compared to other two axes. And all three directional bulk moduli of
\emph{Pnma} ThO$_{2}$ are apparently smaller than that of \emph{Pnma}
PuO$_{2}$ and UO$_{2}$, which indicates relative weaker covalency of ThO$_{2}$
compared with PuO$_{2}$ and UO$_{2}$ in their high pressure phase.

\subsubsection{Debye temperature}

\begin{table}[ptb]
\caption{Calculated density (in g/cm$^{3}$), transverse ($\upsilon_{t}$),
longitudinal ($\upsilon_{l}$) and average ($\upsilon_{m}$) sound velocities
(in m/s) derived from bulk and shear modulus, and Debye temperature (in K) for
actinide dioxides. Results of NpO$_{2}$ and ThO$_{2}$ are calculated by using
the elastic data included in our previous studies \cite{WangNpO2,WangThO2}.}%
\label{Debye}
\begin{ruledtabular}
\begin{tabular}{cccccccccccccc}
Compound&Phase&$\rho$&$\upsilon_{t}$&$\upsilon_{l}$&$\upsilon_{m}$&$\theta_{D}$\\
\hline
PuO$_{2}$&\emph{Fm$\bar{3}$m}&11.892&2477.6&5206.3&2787.0&354.5\\
PuO$_{2}$&\emph{Pnma}&12.812&2387.0&4976.0&2683.9&350.0\\
NpO$_{2}$&\emph{Fm$\bar{3}$m}&11.347&2835.0&5566.5&3176.8&401.2\\
UO$_{2}$&\emph{Fm$\bar{3}$m}&11.084&2841.8&5552.7&3183.4&398.1\\
UO$_{2}$&\emph{Pnma}&11.957&2384.2&4948.5&2680.2&343.7\\
ThO$_{2}$&\emph{Fm$\bar{3}$m}&9.880&2969.1&5575.5&3317.3&402.6\\
ThO$_{2}$&\emph{Pnma}&10.505&2504.6&4738.4&2799.8&346.8\\
\end{tabular}
\end{ruledtabular}
\end{table}

In study of the sound velocities and Debye temperature, our calculated results
of AFM PuO$_{2}$ and AFM UO$_{2}$ in their fluorite phase and high pressure
cotunnite phase are presented in Table \ref{Debye}. For comparison, results of
NpO$_{2}$ and ThO$_{2}$ calculated by using the elastic data included in our
previous studies \cite{WangNpO2,WangThO2} are also listed. As seen from Table
\ref{Debye}, in their \emph{Fm$\bar{3}$m} structure, Debye temperature and
sound velocity of UO$_{2}$, NpO$_{2}$, and ThO$_{2}$ are higher than that of
PuO$_{2}$. This is interestingly associated with the fact that Debye
temperature ($\theta_{D}$) and Vickers hardness (\emph{H}) of materials follow
an empirical relationship \cite{Abrahams}:
\begin{equation}
\theta_{D}\propto\emph{H}^{1/2}\Omega^{1/6}\emph{M}^{-1/2},
\end{equation}
where \emph{M} is molar mass and $\Omega$ is molecular volume. Although
\emph{Fm$\bar{3}$m} PuO$_{2}$ has close value of hardness compared with
UO$_{2}$ and NpO$_{2}$, it has relatively smaller molecular volume and larger
molar mass, as a consequence, has a lower Debye temperature than that of
UO$_{2}$ and NpO$_{2}$. As for the \emph{Pnma} structure, Debye temperature
and sound velocity of PuO$_{2}$, UO$_{2}$, and ThO$_{2}$ are lower than those
of their \emph{Fm$\bar{3}$m} structure. This is since that the \emph{Pnma}
PuO$_{2}$, UO$_{2}$, and ThO$_{2}$ have smaller values of volume, bulk and
shear moduli compared to their ground-state \emph{Fm$\bar{3}$m} structure.

\subsubsection{Hardness}

Hardness is also one fundamental physical quantity when considering the phase
stability and mechanical properties. According to the hardness conditions
\cite{Kaner}, the hardness is closely related to interatomic distances, number
of nearest neighbors, directional bonding, anisotropy, and the indenter
orientation in the structures. To date there is still no available calculation
method involving hardness anisotropy in different dimensions in the
literature. In spite of that, however, recently a semiempirical approach of
hardness was raised by Simunek and Vackar \cite{SimunekPRL,SimunekPRB} in
terms of the atomistic bond strength properties. This approach has been
successfully tested by the authors on the more than 30 binary structures with
zinc blende, cubic fluorite, rock salt crystals, and also for highly covalent
crystals \cite{SimunekPRL,SimunekPRB}. There is no need for all those
high-symmetry structures to consider the anisotropy. Also, this method has
been applied to the crystals involving covalent and ionic bonding characters,
such as in the compounds of transition metal and light atoms, and generalized
to the complex structure with more than two different bond strengths
\cite{SimunekPRL,SimunekPRB}. Moreover, in study of covalent crystals, the
results of the method raised by Simunek and Vackar
\cite{SimunekPRL,SimunekPRB} agree well with those of another hardness method
of Gao and co-workers \cite{GaoPRL}. Therefore, the hardness of optimized
cubic fluorite structures of actinide dioxides can be calculated by the method
of Simunek and Vackar \cite{SimunekPRL,SimunekPRB}. In the case of two atoms 1
and 2 forming one bond of strength \emph{s$_{12}$} in a unit cell of volume
$\Omega$, the expression for hardness has the form \cite{SimunekPRL}
\begin{equation}
H=(C/\Omega)b_{12}s_{12}e^{-\sigma\!f_{2}},
\end{equation}
where
\begin{equation}
s_{12}=\sqrt{(e_{1}e_{2})}/(n_{1}n_{2}d_{12}), e_{i}=Z_{i}/r_{i}%
\end{equation}
and
\begin{equation}
f_{2}=(\frac{e_{1}-e_{2}}{e_{1}+e_{2}})^{2}=1-[\sqrt{(e_{1}e_{2})}%
/(e_{1}+e_{2})]^{2}%
\end{equation}
are the strength and ionicity of the chemical bond, respectively, and
\emph{d$_{12}$} is the interatomic distance; \emph{C}=1550 and $\sigma$=4 are
constants. The radius \emph{r$_{i}$} is chosen to make sure that the sphere
centered at atoms \emph{i} in a crystal contains exactly the valence
electronic charge \emph{Z$_{i}$}. For fluorite structure PuO$_{2}$,
\emph{b$_{12}$}=32 counts the interatomic bonds between atoms Pu (1) and O (2)
in the unit cell, \emph{n$_{1}$}=8 and \emph{n$_{2}$}=4 are coordination
numbers of atom Pu and O, respectively, \emph{r$_{1}$}=1.698 ({\AA }) and
\emph{r$_{2}$}=1.005 ({\AA }) are the atomic radii for Pu and O atoms,
respectively, \emph{Z$_{1}$}=16 and \emph{Z$_{2}$}=6 are valence charge for Pu
and O atoms, respectively, \emph{d$_{12}$}=2.32 ({\AA }) is the interatomic
distance, and $\Omega$=154.16 ({\AA }$^{3}$) is the volume of unit cell. As
for fluorite UO$_{2}$, \emph{b$_{12}$}=32, \emph{n$_{1}$}=8, \emph{n$_{2}$}=4,
\emph{r$_{1}$}=1.737 ({\AA }), \emph{r$_{2}$}=1.003 ({\AA }), \emph{Z$_{1}$%
}=14, \emph{Z$_{2}$}=6, \emph{d$_{12}$}=2.36 ({\AA }), $\Omega$=161.82
({\AA }$^{3}$). Using Eqs. (14)-(16), we obtain \emph{s$_{12}$}=0.1010 and
\emph{f$_{2}$}=0.0503 for PuO$_{2}$ and \emph{s$_{12}$}=0.0919 and
\emph{f$_{2}$}=0.0219 for UO$_{2}$. Therefore, the hardness of PuO$_{2}$ at
its ground-state fluorite structure can be described as \emph{H}=26.6 (GPa)
and for UO$_{2}$ \emph{H}=25.8 (GPa). Clearly, the hardness of PuO$_{2}$,
almost equal to the hardness of NpO$_{2}$ (26.5 GPa) \cite{WangNpO2}, is
slightly larger than that of UO$_{2}$. Besides, these three values of hardness
are all larger than that of ThO$_{2}$ (22.4 GPa) \cite{WangThO2}.

\subsubsection{Theoretical tensile strength}

\begin{figure}[ptb]
\begin{center}
\includegraphics[width=0.8\linewidth]{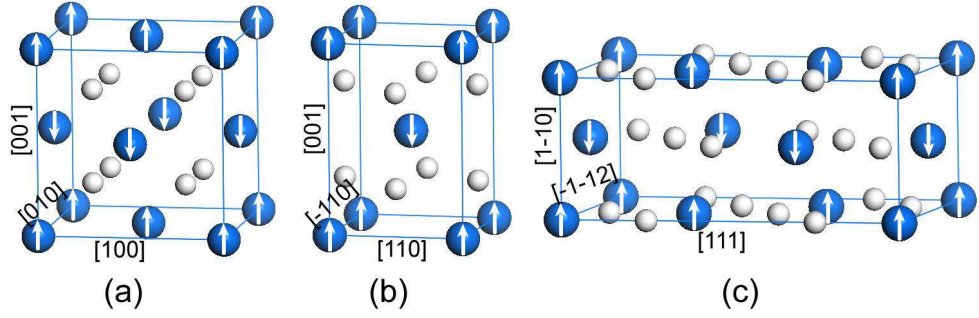}
\end{center}
\caption{(Color online) Schematic illustration of tension along (a) [001], (b)
[110], and (c) [111] orientations. The AFM order is indicated by white arrows
attached on Pu atoms.}%
\label{tensile1}%
\end{figure}

Although many efforts have been paid on PuO$_{2}$, little is known on its
theoretical tensile strength. The ideal strength of materials is the stress
that is required to force deformation or fracture at the elastic instability.
Although the strength of a real crystal can be changed by the existing cracks,
dislocations, grain boundaries, and other microstructural features, its
theoretical value can never be raised, i.e., the theoretical strength sets an
upper bound on the attainable stress. Here, we employ a first-principles
computational tensile test (FPCTT) \cite{Zhang} to calculate the stress-strain
relationship and obtain the ideal tensile strength by deforming the PuO$_{2}$
crystals to failure. The anisotropy of the tensile strength is tested by
pulling the initial fluorite structure along the [001], [110], and [111]
directions. As shown in Fig. \ref{tensile1}, three geometric structures are
constructed to investigate the tensile strengthes in the three typical
crystallographic directions: \ref{tensile1}(a) shows a general fluorite
structure of PuO$_{2}$ with four Pu and eight O atoms; \ref{tensile1}(b) a
body-centered tetragonal (bct) unit cell with two Pu and four O; and
\ref{tensile1}(c) a orthorhombic unit cell with six Pu and twelve O. In FPCTT,
the tensile stress is calculated according to the Nielsen-Martin scheme
\cite{Nielsen} $\sigma_{\alpha\beta}$=$\frac{1}{\Omega}$$\frac{\partial
\newline E_{\mathrm{{total}}}}{\partial\newline\varepsilon_{\alpha\beta}}$
where $\varepsilon_{\alpha\beta}$ is the strain tensor ($\alpha, \beta$=1,2,3)
and $\Omega$ is the volume at the given tensile strain. Tensile process along
the [001], [110], and [111] directions is implemented by increasing the
lattice constants along these three orientations step by step. At each step,
the structure is fully relaxed until all other five stress components vanished
except that in the tensile direction. 

\begin{figure}[ptb]
\begin{center}
\includegraphics[width=1.0\linewidth]{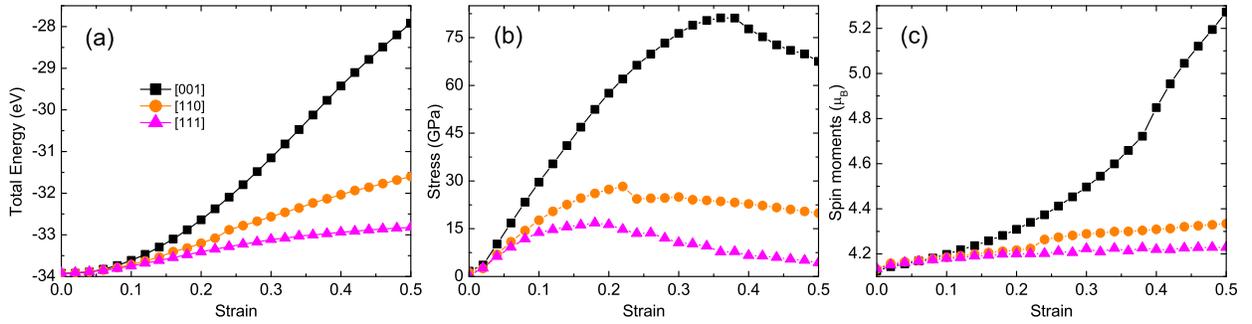}
\end{center}
\caption{(Color online) Dependence of the (a) total energy (per formula unit),
(b) stress, and (c) spin moments on tensile strain for AFM PuO$_{2}$ in the
[001], [110], and [111] directions.}%
\label{tensile2}%
\end{figure}

\begin{table}[ptb]
\caption{Calculated stress maxima and the corresponding strains in the tensile
process.}%
\label{tensile}
\begin{ruledtabular}
\begin{tabular}{ccccc}
Direction&Stress (GPa)&Strain\\
\hline
[001]&81.2&0.36\\
$[110]$&28.3&0.22\\
$[111]$&16.8&0.18\\
\end{tabular}
\end{ruledtabular}
\end{table}

\begin{figure}[ptb]
\begin{center}
\includegraphics[width=1.0\linewidth]{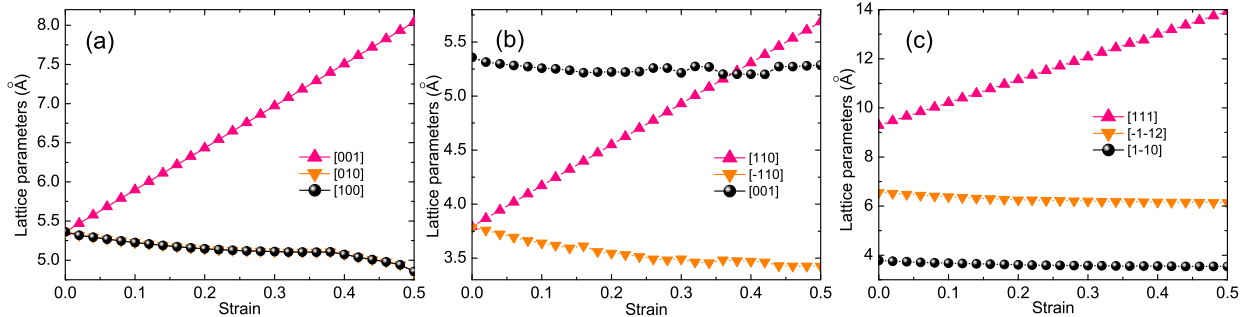}
\end{center}
\caption{(Color online) Dependence of the lattice parameters on tensile strain
for AFM PuO$_{2}$ in the (a) [001], (b) [110], and (c) [111] directions.}%
\label{tensile3}%
\end{figure}

The calculated total energy, stress, and spin moments as functions of uniaxial
tensile strain for AFM PuO$_{2}$ in the [001], [110], and [111] directions are
shown in Fig. \ref{tensile2}. Clearly, all three energy-strain curves increase
with increasing tensile strain, but one can easily find the inflexions by
performing differentiations. Actually, at strains of 0.36, 0.22, and 0.18, the
stresses reach maxima 81.2, 28.3, and 16.8 GPa for pulling in the [001],
[110], and [111] directions, respectively. For clear comparison, all these
maximum stresses (i.e., the theoretical tensile strengthes in the three
typical crystalline orientations) and the corresponding strains are listed in
Table \ref{tensile}. Our results indicate that the [001] direction is the
strongest tensile direction and [111] the weakest. In fact, there are eight
Pu$-$O bonds per formula unit for fluorite PuO$_{2}$. The angle of all eight
bonds with respect to the pulling direction is 45$^{\circ}$ in [001]
direction. However, in [110] direction only four bonds make an angle of
45$^{\circ}$ with the pulling direction. Four other bonds are vertical to the
pulling direction. In [111] direction, two bonds are parallel to the pulling
direction and six others make an angle of about 19.5$^{\circ}$ with the
pulling direction. The bonds vertical to the pulling direction have no
contributions on the tensile strength and the bonds parallel to the pulling
direction are easy to fracture under tensile deformation. Therefore, the fact
that the tensile strength along the [001] direction is stronger than that
along other two directions is understandable. Besides, we note that the stress
in [110] direction experiences an abrupt decrease process after strain up to
0.24. This is due to the fact that the corresponding four Pu$-$O bonds (make
an angle of 45$^{\circ}$ with the pulling direction) have been pulled to
fracture. The fracture behaviors have been clarified by plotting valence
electron charge density maps (not shown). Under the same strain, the abrupt
increase of spin moment can be clearly seen [Fig. \ref{tensile2}(c)]. While
the spin moments in [110] and [111] directions only increase from 4.13 to 4.23
and 4.33 $\mu_{B}$, respectively, the spin moments in [001] direction increase
up to 5.27 $\mu_{B}$ at the end of tensile deformation. In addition, the
evolutions of the lattice parameters with strain in all three tensile
processes are presented in Fig. \ref{tensile3}. One can see from Fig.
\ref{tensile3} that along with the increase of the lattice parameter in the
pulling direction, other two lattice parameters vertical to the pulling
direction are decrease smoothly. In [001] direction, the evolutions of the
lattice parameters along [100] and [010] directions are absolutely same due to
the structural symmetry. For all three tensile deformations, no structural
transition has been observed in our present FPCTT study.

\subsection{Phonon dispersion and thermodynamic properties of fluorite
PuO$_{2}$}

To our knowledge, no experimental phonon frequency results have been published
for PuO$_{2}$. In 2008, Yin and Savrasov \cite{Yin} successfully obtained the
phonon dispersions of both UO$_{2}$ and PuO$_{2}$ by employing the LDA+DMFT
scheme. They found that the dispersive longitudinal optical (LO) modes do not
participate much in the heat transfer due to their large anharmonicity and
only longitudinal acoustic (LA) modes having large phonon group velocities are
efficient heat carriers. In 2009, Minamoto \emph{et al.} \cite{Minamoto}
investigated the thermodynamic properties of PuO$_{2}$ based on their
calculated phonon dispersion within the pure GGA scheme. In present work, we
have calculated the Born effective charges $Z^{\ast}$ and dielectric constants
$\varepsilon_{\infty}$ of PuO$_{2}$ before phonon dispersion calculation due
to their critical importance to correct the LO-TO splitting. For fluorite
PuO$_{2}$, the effective charge tensors for both Pu and O are isotropic
because of their position symmetry. After calculation, the Born effective
charges of Pu and O ions for AFM PuO$_{2}$ are found to be $Z_{\mathrm{{Pu}}%
}^{\ast}$=5.12 and $Z_{\mathrm{{O}}}^{\ast}$=$-$2.56, respectively, within
LDA+\emph{U} formalism with the choice of $U$=$4.0$ eV. In addition, the
macroscopic static dielectric tensor is also isotropic and our computed value
of dielectric constant $\varepsilon_{\infty}$ is 5.95 for the AFM phase. As
for phonon dispersion, the Hellmann-Feynman theorem and the direct method
\cite{Parlinski} are employed to calculate the phonon curves along $\Gamma$%
$-$$X$$-$$K$$-$$\Gamma$$-$$L$$-$$X$$-$$W$$-$$L$ directions in the BZ, together
with the phonon density of states. We use 2$\times$2$\times$2 fcc supercell
containing 96 atoms and 3$\times$3$\times$3 Monkhorst-Pack \emph{k}-point mesh
for the BZ integration. In order to calculate the Hellmann-Feynman forces, we
displace four atoms (two Pu and two O atoms) from their equilibrium positions
and the amplitude of all the displacements is 0.02 \AA .

\begin{figure}[ptb]
\begin{center}
\includegraphics[width=0.5\linewidth]{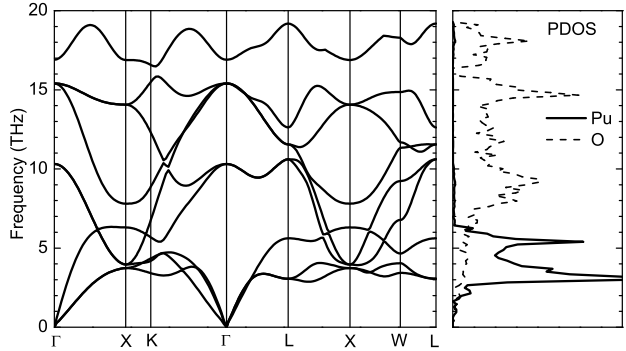}
\end{center}
\caption{Phonon dispersion curves (left panel) and corresponding PDOS (right
panel) for AFM PuO$_{2}$ calculated within LDA+\emph{U} formalism with
\emph{U}=4 eV.}%
\label{phonon}%
\end{figure}

The calculated phonon dispersion curves are displayed in Fig. \ref{phonon} for
the AFM phase of PuO$_{2}$. In the fluorite primitive cell, there are only
three atoms (one Pu and two O atoms). Therefore, nine phonon modes exist in
the dispersion relations. One can see that the LO-TO splitting at $\Gamma$
point becomes evident by the inclusion of polarization effects. In addition,
due to the fact that plutonium atom is heavier than oxygen atom, the vibration
frequency of plutonium atom is lower than that of oxygen atom. As a
consequence, the phonon density of states for PuO$_{2}$ can be viewed as two
parts. One is the part lower than 6.4 THz where the main contribution comes
from the plutonium sublattice, while the other part higher than 6.4 THz are
dominated by the dynamics of the light oxygen atoms.

Thermodynamic properties of AFM PuO$_{2}$ calculated within LDA+\emph{U}
formalism with the choice of $U$=$4.0$ eV are determined by phonon calculation
using the quasiharmonic approximation \cite{Siegel}, under which the Gibbs
free energy \emph{G(T,P)} is written as
\begin{equation}
G(T,P)=F(T,V)+PV.
\end{equation}
Here, \emph{F(T,V)} is the Helmholtz free energy at temperature \emph{T} and
volume \emph{V} and can be expressed as
\begin{equation}
F(T,V)=E(V)+F_{ph}(T,V)+F_{el}(T,V),
\end{equation}
where \emph{E(V)} is the ground-state total energy, \emph{F$_{ph}$(T,V)} is
the phonon free energy and \emph{F$_{el}$(T,V)} is the thermal electronic
contribution. In present study, we focus only on the contribution of atom
vibrations. The \emph{F$_{ph}$(T,V)} can be calculated by
\begin{equation}
F_{ph}(T,V)=k_{B}T\int_{0}^{\infty}g(\omega)\ln\left[  2 \sinh\left(
\frac{\hslash\omega}{2k_{B}T}\right)  \right]  d\omega,
\end{equation}
where $\omega$=$\omega(V)$ represents the volume-dependent phonon frequencies
and $g(\omega)$ is the phonon DOS.

\begin{figure}[ptb]
\begin{center}
\includegraphics[width=0.5\linewidth]{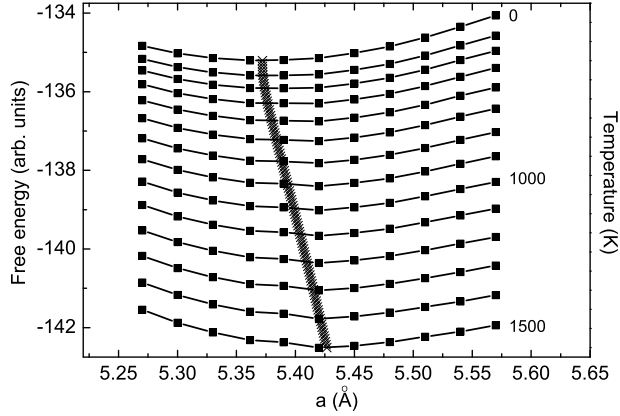}
\end{center}
\caption{Dependence of the free energy \emph{F(T,V)} on crystal lattice
parameter \emph{a} for a number of selected temperatures for AFM PuO$_{2}$
calculated within LDA+\emph{U} formalism with \emph{U}=4 eV.}%
\label{freeE}%
\end{figure}

\begin{figure}[ptb]
\begin{center}
\includegraphics[width=0.5\linewidth]{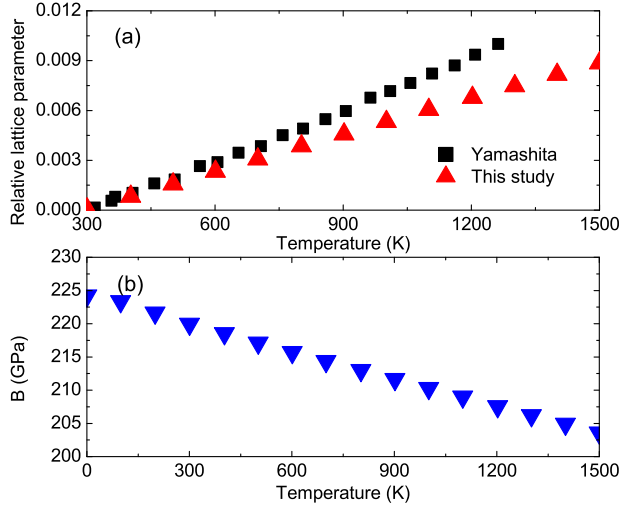}
\end{center}
\caption{(Color online) Temperature dependence of (a) relative lattice
parameter $[a(T)-a(300)]/a(300)$, where a(300) denotes the lattice parameter
at 300 K, and (b) bulk modulus \emph{B(T)} of PuO$_{2}$. Experimental results
from \cite{Yamashita} are also shown in panel (a).}%
\label{latticeT}%
\end{figure}

\begin{figure}[ptb]
\begin{center}
\includegraphics[width=0.5\linewidth]{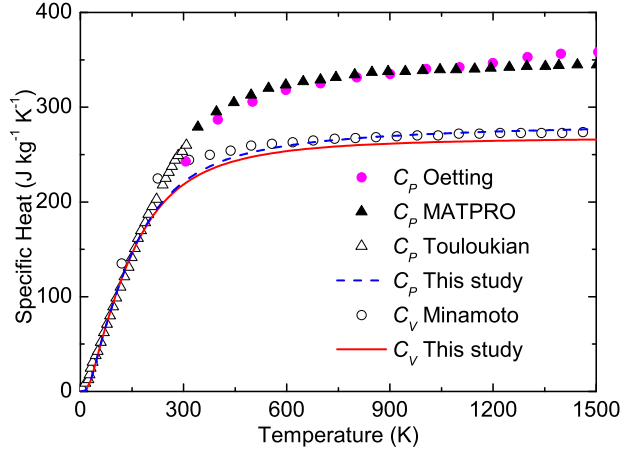}
\end{center}
\caption{(Color online) Calculated specific heat at constant pressure
(\emph{C$_{P}$}) and at constant volume (\emph{C$_{V}$}). For comparison,
previous experimental results from \cite{Oetting,MATPRO,Touloukian,Minamoto}
are also shown.}%
\label{specific}%
\end{figure}

In calculation of $F(T,V)$, the ground-state total energy and phonon
free energy should be calculated by constructing several 2$\times
$2$\times$2 fcc supercells. The temperature \emph{T} appears in
$F(T,V)$ via the phonon term only. Calculated free energy $F(T,V)$
curves of PuO$_{2}$ for temperature ranging from 0 up to 1500 K are
shown in Fig. \ref{freeE}. The locus of the equilibrium lattice
parameters at different temperature \emph{T} are also presented. The
equilibrium volume \emph{V(T)} and the bulk modulus \emph{B(T)} are
obtained by EOS fitting. Figure \ref{latticeT} shows the temperature
dependence of the relative lattice parameter and the bulk modulus.
Experimental results \cite{Yamashita} are also plotted. Clearly, the
lattice parameter increases approximately in the same ratio as that
in experiment \cite{Yamashita}. The bulk modulus \emph{B(T)} is
predicted to decrease along with the increase of temperature. This
kind of temperature effect is very common for compounds and metals.
Besides, the specific heat at constant volume $C_{V}$ can be
directly calculated through
\begin{equation}
C_{V}=k_{B}\int_{0}^{\infty}g(\omega)\left(  \frac{\hslash\omega}{k_{B}%
T}\right)  ^{2}\frac{\exp\frac{\hslash\omega}{k_{B}T}}{(\exp\frac
{\hslash\omega}{k_{B}T}-1)^{2}}d\omega,
\end{equation}
while the specific heat at constant pressure $C_{P}$ can be evaluated by the
thermodynamic relationship $C_{P}-C_{V}=\alpha_{V}^{2}(T)B(T)V(T)T$, where the
isobaric thermal expansion coefficient can be calculated according to the
formula $\alpha_{V}(T)=\frac{1}{V}\left(  \frac{\partial V}{\partial
T}\right)  _{P}$. Calculated $C_{P}$ and $C_{V}$ are presented in Fig.
\ref{specific}. Clearly, the calculated $C_{V}$ is in good agreement with
experiment \cite{Minamoto} in all investigating temperature domain, while the
predicted $C_{P}$ only accords well with the corresponding experimental data
\cite{Touloukian} below 300 K due to the intrinsic fact that near zero
temperature $C_{P}$ and $C_{V}$ approach to the same value. The disagreement
of $C_{P}$ between theory and experiments \cite{Oetting,MATPRO} in high
temperature domain is believed to mainly originate from the quasiharmonic
approximation we use.

\subsection{High pressure behavior of PuO$_{2}$}

\begin{figure}[ptb]
\begin{center}
\includegraphics[width=0.5\linewidth]{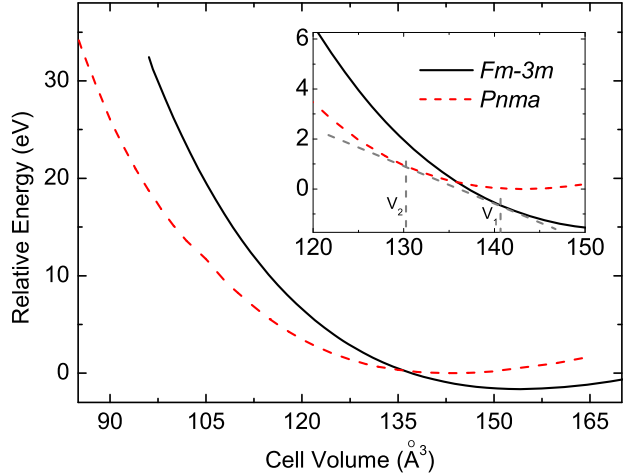}
\end{center}
\caption{(Color online) Comparison of relative energy vs the cell volume for
AFM PuO$_{2}$ in \emph{Fm$\bar{3}$m} and \emph{Pnma} phases. The total energy
of \emph{Pnma} phase at 0 GPa is set as zero. A phase transition at 24.3 GPa
is predicted by the slope of the common tangent rule, as shown in the inset.}%
\label{energy}%
\end{figure}

In the following, we will focus our attention on the behavior of plutonium
dioxide under hydrostatic compression. Two experimentally established
structures, \emph{Fm$\bar{3}$m} and \emph{Pnma} phases, are investigated in
detail. The relative energies (per unit cell) of the two phases at different
volumes are calculated and shown in Fig. \ref{energy}. Obviously, the
\emph{Fm$\bar{3}$m} phase is stable under ambient pressure while under high
pressure the \emph{Pnma} phase becomes stable. According to the rule of common
tangent of two energy curves, a phase transition at 24.3 GPa is predicted by
the slope shown in the inset of Fig. \ref{energy}. Besides, we also determine
the phase transition pressure by comparing the Gibbs free energy as a function
of pressure. At 0 K, the Gibbs free energy is equal to enthalpy \emph{H},
expressed as \emph{H}=\emph{E}+\emph{PV}. The crossing between the two
enthalpy curves (not shown) also gives phase transition pressure of 24.3 GPa,
which is fully consistent with above result in terms of the common tangent
rule. This value is smaller by $\mathtt{\sim}$15 GPa when compared to the
experiment measurement by Dancausse \emph{et al} \cite{Dancausse}. We notice
that Li \emph{et al} \cite{Li} predicted a transition pressure of about 45 GPa
employing the full-potential linear-muffin-tin-orbital (FPLMTO) method.
However, they only considered the NM and FM phases in their calculations of
the \emph{Fm$\bar{3}$m} and \emph{Pnma} PuO$_{2}$. No results were calculated
for the AFM phase. In our present study, we first optimize the \emph{Pnma}
PuO$_{2}$ at constant volume (about 125.2 {\AA }$^{3}$) of the experimental
value at 39 GPa \cite{Dancausse}. Our calculated structural parameters of the
NM, FM, and AFM phases are tabulated in Table \ref{pnma}. Obviously, the AFM
phase is the most stable phase and its structural parameters are well
consistent with experiment. As seen in Table \ref{pnma}, results of AFM phase
at 49 GPa is also consistent with experiment. Therefore, we will only consider
the results of AFM phase in the following discussion. In addition, obvious
discrepancy between two works \cite{Idiri,Dancausse} from the same
experimental group involving the transition pressure of ThO$_{2}$ was noticed
in our previous study \cite{WangThO2}. While the transition was reported
firstly to start at 40 GPa \cite{Dancausse}, later, through improving
experimental measurement technique, which can capture the two phases
cohabitation zone during transition process, Idiri \emph{et al} \cite{Idiri}
observed that the transition really begins around 33 GPa. Our previous study
\cite{WangThO2} predicted that the phase transition of ThO$_{2}$ started at
around 26.5 GPa. Idiri \emph{et al} \cite{Idiri} also stated that the bulk
modulus of PuO$_{2}$ were largely overestimated in their previous work
\cite{Dancausse}. Thus, we believe that their former report \cite{Dancausse}
also overestimated the transition pressure of PuO$_{2}$ in some degree. So the
predicted transition pressure of PuO$_{2}$, which is very close to that of
ThO$_{2}$ \cite{WangThO2}, is understandable.

\begin{table}[ptb]
\caption{Calculated structural parameters (in ${\mathring{A}}$), pressure (in
GPa), and relative energy (in eV) with respect to the total energy at 0 GPa of
the \emph{Pnma} PuO$_{2}$ at two constant volumes from experiment
\cite{Dancausse}. For comparison, the experimental values are also listed.}%
\label{pnma}
\begin{ruledtabular}
\begin{tabular}{cccccccccccccc}
Magnetism&a&b&c&Pressure&Relative Energy\\
\hline
NM&6.502&3.165&6.087&82.9&32.87\\
FM&7.478&3.088&5.424&124.2&31.80\\
AFM&5.585&3.410&6.577&38.6&1.90\\
Expt.&5.64&3.38&6.57&39&\\
Expt.&5.62&3.44&6.49&49&\\
AFM&5.492&3.398&6.539&48.8&2.80\\
\end{tabular}
\end{ruledtabular}
\end{table}

\begin{figure}[ptb]
\begin{center}
\includegraphics[width=0.5\linewidth]{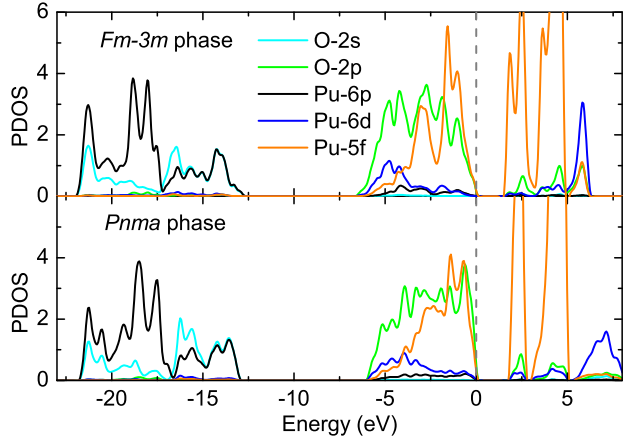}
\end{center}
\caption{(Color online) Partial density of states (PDOS) for (a)
\emph{Fm$\bar{3}$m} phase and (b) \emph{Pnma} phase at around 25 GPa. The
Fermi energy level is zero. The energy gaps of \emph{Fm$\bar{3}$m} phase and
\emph{Pnma} phase are 1.51 eV and 1.65 eV, respectively.}%
\label{twoDOS}%
\end{figure}

Figure \ref{twoDOS} compares the PDOS of the \emph{Fm$\bar{3}$m} and
\emph{Pnma} phases of PuO$_{2}$ at a pressure of around 25 GPa, close to the
transition pressure. One can see evident increase of the band gap from 1.51 eV
in the fluorite phase to 1.65 eV in the cotunnite phase. In study of UO$_{2}$,
Geng \emph{et al} \cite{Geng} also predicted this similar behavior. However,
our previous study of ThO$_{2}$ \cite{WangThO2} found that the band gap is
reduced from \emph{Fm$\bar{3}$m} phase to \emph{Pnma} phase. The reason is
simply that ThO$_{2}$ is a charge-transfer insulator, while PuO$_{2}$ and
UO$_{2}$ are the Mott-type insulators. From Fig \ref{twoDOS}, one can see that
while O 2$s$ and Pu 6$p$ states expand in the low bands, O 2$p$ and Pu
5$f$/6$d$ states are mainly featured near the Fermi level and have prominent
hybridization characters in a width of 6.3 (5.9) eV for \emph{Fm$\bar{3}$m}
(\emph{Pnma}) phase. There is no evident difference between the two phases in
their 5$f$ electronic localization behavior. Therefore, our study of PuO$_{2}$
supports our previous \cite{WangThO2} viewpoint: the phenomenon of volume
collapse during high-pressure phase transition of the actinide dioxides is
mainly attributed to the ionic (instead of electronic) response to the
external compression.

\begin{figure}[ptb]
\begin{center}
\includegraphics[width=0.5\linewidth]{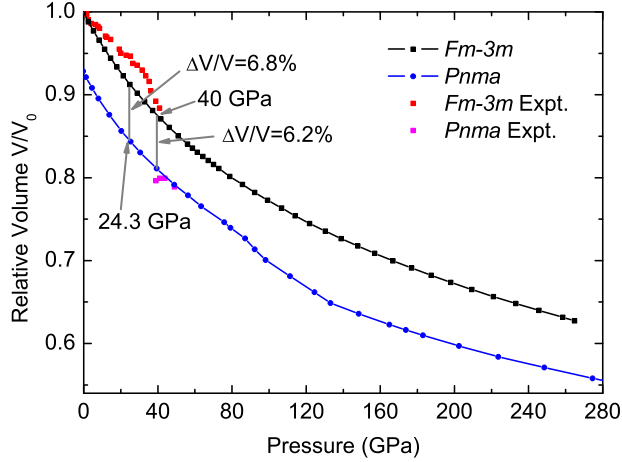}
\end{center}
\caption{(Color online) Calculated compression curves of PuO$_{2}$ compared
with experimental measurements. The volume collapses at our predicted phase
transition point 24.3 GPa and experimental \cite{Dancausse} phase transition
pressure 40 GPa are labeled.}%
\label{pressure}%
\end{figure}

\begin{figure}[ptb]
\begin{center}
\includegraphics[width=0.5\linewidth]{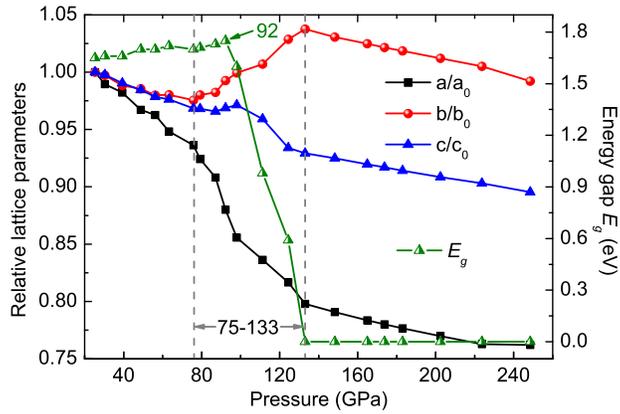}
\end{center}
\caption{(Color online) Pressure behavior of the relative lattice parameters
of the \emph{Pnma} phase, where the drastic change in the relative lattice
constants (region between dashed lines) indicates an isostructural transition.
Besides, the pressure behavior of the insulating band gap is also shown.}%
\label{lattice}%
\end{figure}

\begin{figure}[ptb]
\begin{center}
\includegraphics[width=1.0\linewidth]{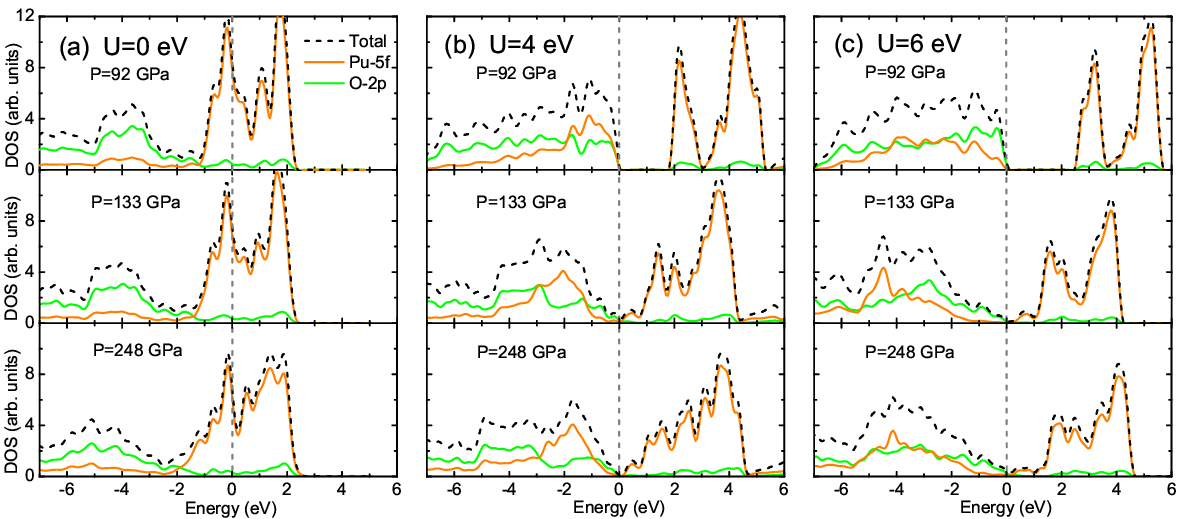}
\end{center}
\caption{(Color online) The total DOS for the cotunnite PuO$_{2}$ AFM phase
calculated at selected pressures within LDA+\emph{U} formalism with (a)
\emph{U}=0 eV, (b) \emph{U}=4 eV, and (c) \emph{U}=6 eV. The projected DOSs
for the Pu 5\emph{f} and O 2\emph{p} orbitals are also shown. The Fermi energy
level is set at zero.}%
\label{metallicDOS}%
\end{figure}

The relative volume \emph{V/V$_{0}$} evolution with pressure for PuO$_{2}$ in
both \emph{Fm$\bar{3}$m} and \emph{Pnma} phases are shown in Fig
\ref{pressure}. For comparison, the experimental \cite{Dancausse} data are
also shown in the figure. Clearly, our calculated P-V equation of state is
well consistent with the experimental measurement for the two phases of
PuO$_{2}$. Specially, at the calculated transition pressure (24.3 GPa), our
result in Fig. \ref{pressure} gives that the volume collapse upon phase
transition is 6.8\%. This value is evidently underestimated compared with the
experimental data of 12\% \cite{Dancausse}. The discrepancy between experiment
and our calculation needs more experimental and theoretical works to examine.
After phase transition, we also find an isostructural transition occurring
between 75 and 133 GPa for the \emph{Pnma} phase. This isostructural
transition of actinide dioxides was first found in DFT calculations of
UO$_{2}$ by Geng \emph{et al} \cite{Geng} and then observed in study of
ThO$_{2}$ \cite{WangThO2}. The pressure dependence of the three lattice
parameters (with respect to their values at calculated transition pressure
24.3 GPa) for the \emph{Pnma} phase of PuO$_{2}$ are plotted in Fig.
\ref{lattice}. Similar to the studies of UO$_{2}$ and ThO$_{2}$
\cite{Geng,WangThO2}, in pressure region before 75 GPa, the responses of the
three relative lattice parameters to the compression are anisotropic in the
sense that the compression of the middle axis $a$ is most rapid compared to
those of the long axis $c$ and small axis $b$, which vary upon compression
almost in the same tendency. When the pressure becomes higher to be between 75
and 133 GPa, remarkably, it reveals in Fig. \ref{lattice} that all the three
relative lattice parameters undergo dramatic variations by the fact that the
small axis $b$ has a strong rebound and the middle $a$ is collapsed. When the
pressure is beyond 133 GPa, then the variations of the three relative lattice
parameters become smooth and approach isotropic compression. This signifies a
typical isostructural transition for the \emph{Pnma} phase of PuO$_{2}$.

Moreover, we also present in Fig. \ref{lattice} the evolution of the
insulating band gap with pressure for \emph{Pnma} phase of
PuO$_{2}$. Apparently, the band gap behaves smooth in pressure
region of 25-92 GPa, then turns to decrease suddenly from about 1.75
eV to zero under compression from 92 to 133 GPa. This clearly
indicates that the \emph{Pnma} phase will occur a metallic
transition after external pressure exceeds 133 GPa. As Fig.
\ref{metallicDOS}(b) shows, with increasing pressure in the
crossover range between 92 and 133 GPa, the 5$f$ electrons in the
cotunnite phase of PuO$_{2}$ are more delocalized and the 5$f$ bands
are largely broadened. As a result, the Mott-type band gap is
narrowed and even blurred, which is characterized by finite
occupancies of O 2$p$ and Pu 5$f$ orbitals at the Fermi level, by
the increasing kinetic energy of 5$f$ electrons. In order to see the
pressure behavior of electronic structure with different values of
Hubbard parameter \emph{U} for cotunnite PuO$_{2}$, we have plotted
in Fig. 15 the total DOS and PDOS for the cotunnite PuO$_{2}$ AFM
phase calculated at 92, 133, and 248 GPa within LDA+\emph{U}
formalism with \emph{U}=0, 4, and 6 eV. It clearly shows that (i)
the pure LDA always produces incorrect electronic structure in the
full pressure region we considered, and (ii) it is not imperative to
adjust $U$ as varying pressure by the fact revealed in Figs. 15(b)
and 15(c) that the electronic properties and insulator-metal
transition behavior at high pressures show the similar character for
the two choices of Hubbard parameter $U$=$4$ eV and $U$=$6$ eV.
Based on this observation, in the above high-pressure calculations
we have fixed the value of $U$ to be $4$ eV. It should be stressed
that the metallic transition is not unique for PuO$_{2}$. Similar
phenomenon has also been observed in other actinide dioxides
\cite{Geng}. Besides, the variation of the local magnetic moment of
plutonium atoms is almost same for \emph{Fm$\bar{3}$m} and
\emph{Pnma} phases, implying that the magnetic property is
insensitive to the structure transition in PuO$_{2}$. Actually, the
calculated amplitude of local spin moment varies from $\sim$4.1 to
$\sim$3.8 $\mu_{B}$ (per Pu atom) for both fluorite and cotunnite
phases in pressure range from 0 to 255 GPa. No paramagnetic
transition for this material has been observed in present study.

\section{CONCLUSIONS}

In conclusion, the ground state properties as well as the high pressure
behavior of PuO$_{2}$ were studied by means of the first-principles
DFT+\emph{U} method. By choosing the Hubbard \emph{U} parameter around 4 eV
within the LDA+\emph{U} approach, the electronic structure, lattice
parameters, and bulk modulus were calculated for both the ambient
\emph{Fm$\bar{3}$m} and the high-pressure \emph{Pnma} phases of PuO$_{2}$ and
were shown to accord well with experiments. Results for UO$_{2}$ were also
presented for comparison. Based on these results, the Pu$-$O and U$-$O bonds
were interpreted as displaying a mixed ionic/covalent character by electronic
structure analysis. After comparing with our previous calculations of
NpO$_{2}$ and ThO$_{2}$, we demonstrated that the Pu$-$O, U$-$O, and Np$-$O
bonds have stronger covalency than the Th$-$O bond. The ionicity of Th$-$O
bond was found to be the largest among these four kinds of actinide dioxides.
In addition, the stability of the two phases at zero pressure was predicted
through calculating elastic constants and phonon dispersion. The hardness,
Debye temperature, ideal tensile strength, and thermodynamic properties were
calculated and analyzed to support the practical application of PuO$_{2}$. We
showed that the hardness of \emph{Fm$\bar{3}$m} phase is $\mathtt{\sim}$27 GPa
and the Debye temperatures of \emph{Fm$\bar{3}$m} and \emph{Pnma} phases are
354.5 and 350.0 K, respectively. For \emph{Fm$\bar{3}$m} PuO$_{2}$, the ideal
tensile strengths are calculated within FPCTT to be 81.2, 28.3, and 16.8 GPa
in tensile deformations along the [100], [110], and [111] directions,
respectively. The volume thermal expansion and specific heat at constant
volume curves are well consistent with available experiments. However, the
discrepancy between measured and our calculated specific heat at constant
pressure in high temperature domain is evident. This needs further theoretical
work by including the anharmonic ionic contribution to decrease this kind of discrepancy.

In studying the pressure behavior of PuO$_{2}$, we showed that the
\emph{Fm$\bar{3}$m}$\mathtt{\rightarrow}$\emph{Pnma} transition occurs at 24.3
GPa. Although this value is large smaller than the experimental report, we
believe that our calculated result is reasonable. One reason is that the
lattice parameters of \emph{Pnma} PuO$_{2}$ AFM phase calculated at around 39
and 49 GPa are precisely consistent with experiment. Another is the fact that
the experiment needs improvement as having been indicated in study of UO$_{2}$
and ThO$_{2}$ \cite{Idiri}. Furthermore, we extended the pressure up to 280
GPa for the two structures of PuO$_{2}$. A metallic transition at around 133
GPa and an isostructural transition in pressure range of 75-133 GPa were
predicted for the \emph{Pnma} phase. Also, the calculated amplitude of local
spin moment only varies from $\sim$4.1 to $\sim$3.8 $\mu_{B}$ (per Pu atom)
for both fluorite and cotunnite phases in pressure range from 0 to 255 GPa. No
paramagnetic transition for this material has been observed.

\begin{acknowledgments}
We gratefully thank G. H. Lu, H. B. Zhou, and X. C. Li for illustrating
discussion on FPCTT. This work was supported by the Foundations for
Development of Science and Technology of China Academy of Engineering Physics
under Grant No. 2009B0301037.
\end{acknowledgments}

\end{document}